\documentstyle[12pt,epsfig]{article}

\newcommand{\half}{{\textstyle\frac{1}{2}}}
\newcommand{\ee}{e^+e^-}
\newcommand{\as}{\alpha_s}
\renewcommand{\ae}{\alpha_{\mbox{\scriptsize eff}}}
\newcommand{\asPT}{\alpha_s^{\mbox{\scriptsize PT}}}
\newcommand{\eps}{\epsilon}
\newcommand{\beq}{\begin{equation}}
\newcommand{\eeq}{\end{equation}}
\newcommand{\cl}[1]{{\cal #1}}
\newcommand{\rf}[1]{(\ref{#1})}
\newcommand{\sect}[1]{\section{#1}\setcounter{equation}{0}}

\newcommand{\prlett}[1]{{\it Phys.~Rev.~Lett.~\bf #1}}
\newcommand{\nphysb}[1]{{\it Nucl.~Phys.~\bf B #1}}
\newcommand{\plettb}[1]{{\it Phys.~Lett.~\bf B #1}}
\newcommand{\phrevd}[1]{{\it Phys.~Rev.~\bf D #1}}
\newcommand{\jinrrc}[1]{{\it JINR~Rapid~Comm.~\bf #1}}
\newcommand{\npproc}[1]{{\it Nucl.~Phys.~Proc.~Suppl.~\bf #1}}
\newcommand{\jhephy}[1]{{\it JHEP~\bf #1}}
\newcommand{\ephysj}[1]{{\it Eur.~Phys.~J.~\bf #1}}
\newcommand{\zphysc}[1]{{\it Z.~Phys.~\bf C #1}}
\newcommand{\sovphy}[1]{{\it Sov.~Phys.~JETP~\bf #1}}
\newcommand{\mpleta}[1]{{\it Mod.~Phys.~Lett.~\bf A #1}}
\newcommand{\ijmpha}[1]{{\it Int.~Jour.~Mod.~Phys.~\bf A #1}}

\begin{document}
\begin{titlepage}
\begin{flushright}
Cavendish-HEP-98/14\\
hep-ph/9810292\\
October 1998
\end{flushright}              
\vspace*{\fill}
\begin{center}
{\Large \bf Power Corrections to Flavour-Singlet\\[1ex]
Structure Functions\footnote{Research supported by the U.K. Particle
Physics and Astronomy Research Council.}}
\end{center}
\par \vskip 5mm
\begin{center}
        G.E.~Smye\\
        Cavendish Laboratory, University of Cambridge,\\
        Madingley Road, Cambridge CB3 0HE, U.K.\\
\end{center}
\par \vskip 2mm
\begin{center} {\large \bf Abstract} \end{center}
\begin{quote}
We investigate the power-suppressed corrections to
structure functions in flavour singlet deep inelastic
lepton scattering, to complement the previous results
for the non-singlet contribution. Our method is a dispersive
approach based on an analysis of Feynman graphs containing
massive gluons; and our results agree with those obtained from leading 
infrared renormalon contributions. As in non-singlet deep inelastic scattering
we find that the leading corrections are proportional to $1/Q^2$. We find
that the singlet contribution becomes important at small $x$.
\end{quote}
\vspace*{\fill}
\end{titlepage}

\sect{Introduction}
The study of structure functions in deep inelastic lepton
scattering (DIS) has received a great impetus from the increasing
quantity and kinematic range of the HERA data \cite{expstrfns}. Although
these functions cannot be calculated using perturbative QCD,
their asymptotic scaling violations (logarithmic $Q^2$ dependence)
can be predicted and used to measure the strong coupling $\as$. (The
same is true of the study of final-state properties, such as fragmentation
functions and event shape variables.)

One problem with the measurement of $\as$ using scaling violation,
in either structure or fragmentation functions, is that there
is $Q^2$ dependence associated with power-suppressed (higher-twist)
contributions, in addition to the dominant logarithmic dependence.
These contributions need to be estimated in order to make use
of the wide $Q^2$ coverage of HERA.

Recently so-called `renormalon' or `dispersive' methods of
estimating power-suppressed terms have been applied to a wide
variety of observables. By looking at the behaviour of
the QCD perturbation series in high orders, one can identify
unsummable, factorially divergent sets of contributions
(infrared renormalons \cite{renormalons,ren2,ren3}) which indicate
that non-perturbative power-suppressed corrections must
be included. The $Q^2$-dependence of the leading correction
to a given quantity can be inferred, and by making further
universality assumptions one may also estimate its magnitude.
Tests of these ideas provide information on the transition
from the perturbative to the non-perturbative regime in QCD.
In particular, one can investigate the possibility that an
approximately universal low-energy form of the strong coupling
may be a useful phenomenological concept \cite{dmw,aspapers,asmodel}.

Such an approach has been applied with some success to $\ee$
fragmentation functions \cite{eefrag} and event shape variables
\cite{webber,eevent,dweevent},
and to flavour non-singlet DIS structure functions \cite{dmw,dwdisstr,disstr},
fragmentation functions \cite{disfrg} and event shape variables \cite{disev}.
Note that special care has to be taken when considering non-inclusive
observables such as fragmentation functions and event shape variables
\cite{milan}. There are also renormalon model results for photon-photon
scattering \cite{hautmann}
and structure functions in flavour singlet DIS \cite{stein}.

In the present paper we perform the singlet structure function
calculation using the dispersive method. We find agreement with the
renormalon model result: as in non-singlet DIS, the predicted leading
power corrections to these quantities are proportional
to $1/Q^2$, but their functional forms are different. In particular there
are contributions proportional to $(\log Q^2)/Q^2$, which are not found
in the non-singlet case. The hypothesis that
they are related to a universal low-energy strong
coupling implies that their magnitudes
are given by a universal non-perturbative parameters.

In the following section we review the approach of reference \cite{dmw}.
Section \ref{strucfns} presents the standard leading-order
perturbative treatment of DIS structure functions. In section \ref{calc}
we estimate the power-suppressed corrections using the method
outlined in section \ref{dispsec}. Our results are
summarized briefly in section \ref{concl}.

\sect{The Dispersive Approach to Power Corrections}
\label{dispsec}
We assume that the QCD running coupling $\alpha_s(k^2)$ can be defined for all positive $k^2$, and that apart from a branch cut along the negative real axis there are no singularities in the complex plane. It follows that we may write the dispersion relation:
\beq
\label{disprel}
\as(k^2) = - \int_0^\infty\frac{d\mu^2}{\mu^2+k^2}\rho_s(\mu^2)\; ,
\eeq
where the `spectral function' $\rho_s$ represents the discontinuity across the cut:
\beq
\rho_s(\mu^2)=\frac{1}{2\pi i}\biggl\{\as(\mu^2 e^{i\pi})-\as(\mu^2 e^{-i\pi})\biggr\} = \frac{1}{2\pi i}\mbox{Disc }\as(-\mu^2)\; .
\eeq

We now consider the calculation of some observable $F$ in an improved one-loop approximation which takes into account one-gluon contributions plus those higher-order terms that lead to the running of $\as$. As discussed in \cite{dmw}, we expect, for squared Feynman diagrams containing a single gluon, that
\beq
F = \as(0)\cl{F}(0) + \int_0^\infty \frac{d\mu^2}{\mu^2}\,\rho_s(\mu^2)\,\cl{F}(\mu^2/Q^2)\;,
\eeq
where the {\it characteristic function} $\cl{F}(\mu^2/Q^2)$ is obtained by one-loop evaluation of $F$ (divided by $\as$) with the gluon mass set equal to $\mu$ \cite{ren2,webber}. The first term on the right-hand side represents the contributions in which a single gluon is produced or exchanged, while the second represents those with more complex final or virtual states (e.g. the `decay products' of a virtual gluon, which contribute to the running of $\as$).

We can eliminate $\as(0)$ by means of the dispersion relation \rf{disprel}:
\beq
\label{Feq}
F = \int_0^\infty \frac{d\mu^2}{\mu^2}\,\rho_s(\mu^2)\,\left[\cl{F}(\mu^2/Q^2)-\cl{F}(0)\right]\;.
\eeq

By rotating the integration contour separately in the two terms of the discontinuity, we obtain the following contribution to the observable $F$:
\beq
F = \int_0^\infty \frac{d\mu^2}{\mu^2}\,\as(\mu^2)\,\cl{G}(\mu^2/Q^2) \;+\; I_\infty[\alpha_s]\;,
\eeq
where, setting $\mu^2/Q^2 = \eps$,
\beq
\label{Gdef}
\cl{G}(\eps) = -\frac{1}{2\pi i}\mbox{Disc }\cl{F}(-\eps)\;,
\eeq
and the functional $I_\infty$ is some integral around a contour at infinity.

Non-perturbative effects at long distances are expected to give rise to a modification in the strong coupling at low scales, $\delta\as(\mu^2) = \as(\mu^2) - \asPT(\mu^2)$, $\asPT(\mu^2)$ being the perturbatively-calculated running coupling. This generates the following non-perturbative correction to the perturbative prediction:
\beq
\label{deltaF}
\delta F = \int_0^\infty \frac{d\mu^2}{\mu^2}\,\delta\as(\mu^2)\,\cl{G}(\mu^2/Q^2)\;.
\eeq

Since $\delta\as(\mu^2)$ is limited to low values of $\mu^2$, there is no contribution to $\delta F$ from the contour at infinity. Furthermore, the asymptotic behaviour of $\delta F$ at large $Q^2$ is controlled by the behaviour of $\cl{F}(\eps)$ as $\eps\to 0$.  We see from \rf{Gdef} that no terms analytic at $\eps=0$ can contribute to $\delta F$.  On the other hand non-analytic terms at small $\eps$ do contribute, the relevant terms being:
\beq
\cl{F}\sim a_1 \frac{C_F}{2\pi}\sqrt{\eps}
\qquad\Longrightarrow\qquad
\delta F = -\frac{a_1}{\pi}\frac{\cl{A}_1}{Q}\;,
\eeq
and
\beq
\cl{F}\sim a_2 \frac{C_F}{2\pi}\eps\log\eps
\qquad\Longrightarrow\qquad
\delta F = a_2\frac{\cl{A}_2}{Q^2}\;,
\eeq
where
\beq
\cl{A}_q \equiv \frac{C_F}{2\pi}\int_0^\infty \frac{d\mu^2}{\mu^2}\,\mu^q\,\delta\as(\mu^2)\;.
\eeq

Notice that we express the result \rf{deltaF} directly in
terms of the modification to the strong coupling itself,
rather than that in the derived quantity $\ae$ which was used
in some previous publications \cite{dmw,dweevent,dwdisstr},
\beq
\ae(\mu^2) =
\frac{\sin(\pi\mu^2\,d/d\mu^2)}{\pi\mu^2\,d/d\mu^2}\as(\mu^2)
= \as(\mu^2)-\frac{\pi^2}{6}\left(\mu^2\frac{d}{d\mu^2}\right)^2\as(\mu^2)
+\ldots\;.
\eeq
Although $\as$ and $\ae$ are similar in the perturbative region,
they differ substantially at low scales, and the former probably
has a simpler behaviour. For example, the even moments of the
effective coupling modification,
\beq
A_{2p} \equiv \frac{C_F}{2\pi}
\int_0^\infty \frac{d\mu^2}{\mu^2}\,\mu^{2p}\,\delta\ae(\mu^2)\;,
\eeq
have to vanish for all integer values of $p$, whereas those of
$\delta\as$ do not. The translation dictionary for the moments
is in any case rather simple:
\beq
\cl{A}_{2p+1} = (-1)^p\,(p+\half)\pi\,A_{2p+1}\;,\;\;\;\;\;
\cl{A}_{2p}   = (-1)^p\,p\,A'_{2p}\;,
\eeq 
where
\beq
A'_{2p} \equiv \frac{d}{dp}A_{2p} = \frac{C_F}{2\pi}
\int_0^\infty \frac{d\mu^2}{\mu^2}\,\mu^{2p}\,\log\mu^2\,
\delta\ae(\mu^2)\;.
\eeq

Studies of the non-singlet contribution to DIS structure functions suggest that $\cl{A}_2 = -A'_2 \simeq 0.2 \mbox{GeV}^2$ \cite{dwdisstr}.

This is the formulation of the dispersive approach to power behaved corrections used where there is a single gluon propagator. However, the calculation which follows in section \ref{calc} involves two gluons, and so the above argument needs to be generalised.

Where a squared Feynman diagram contains two or more gluons, the generalisation of the above requires a separate independent dispersion relation and dispersive variable associated with each gluon. This leads to a characteristic function $\cl{F}(\eps_1,\cdots,\eps_n)$ in all of these variables, and, following the above argument, we will require those terms of $\cl{F}$ which are non-analytic at 0 in all the arguments. However, in the case where there are two or more internal gluons all constrained to have the same 4-momentum, we can simplify this to require only one dispersive variable for all of them, as discussed below.

Suppose a squared Feynman diagram contains $n$ such gluons whose 4-momenta are all constrained to be equal to $k$, for some $k$. (Although only the case $n=2$ is relevant to phenomenology, its treatment takes no fewer lines than that of the general case.) Then, by defining $\rho = - k^2 / Q^2$, we see that the dependence of the characteristic function $\cl{F}$ on $\eps_1,\cdots,\eps_n$ is given by
\beq
\cl{F}(\eps_1,\cdots,\eps_n) = \int\frac{d\rho\,f(\rho)}{(\rho+\eps_1)\cdots(\rho+\eps_n)}\;,
\eeq
where the integration limits and the function $f$ depend on the particular calculation. This may be expressed in partial fractions in the form
\beq
\cl{F}(\eps_1,\cdots,\eps_n) = \sum_{i=1}^n\biggl[\prod_{j\ne i}\frac{1}{\eps_j-\eps_i}\biggr]\int\frac{d\rho\,f(\rho)}{\rho+\eps_i}\;,
\eeq
which yields the result
\beq
\label{Fdecomp}
\cl{F}(\eps_1,\cdots,\eps_n) = \sum_{i=1}^n\biggl[\prod_{j\ne i}\frac{-\eps_i}{\eps_j-\eps_i}\biggr]\hat{\cl{F}}(\eps_i)\;,
\eeq
where
\beq
(-\eps)^{n-1}\hat{\cl{F}}(\eps) = \int\frac{d\rho\,f(\rho)}{\rho+\eps}\;.
\eeq

Now the generalisation of the one-gluon result \rf{Feq} to the multiple-gluon case (which can easily be seen from a generalisation of the argument in \cite{dmw}) is
\beq
F = (-1)^n\prod_{j=1}^n\biggl[\int_0^\infty\frac{d\mu_j^2}{\mu_j^2}\rho_s(\mu_j^2)\sum_{i_j=0}^1(-1)^{i_j}\biggr]{\cal F}(i_1\mu_1^2/Q^2,\cdots,i_n\mu_n^2/Q^2)\;.
\eeq
Substituting \rf{Fdecomp} into this gives
\beq
F = (-1)^n\prod_{j=1}^n\biggl[\int_0^\infty\frac{d\mu_j^2}{\mu_j^2}\rho_s(\mu_j^2)\sum_{i_j=0}^1(-1)^{i_j}\biggr]\sum_{k=1}^n\biggl[\prod_{l\ne k}\frac{-i_k\mu_k^2}{i_l\mu_l^2-i_k\mu_k^2}\biggr]\hat{\cal F}(i_k\mu_k^2/Q^2)\;.
\eeq
Rotating the integration contour over $\mu_1^2$, as in the single-gluon case, and noting that only terms with $k=1$ and $i_1=1$ contribute, we obtain
\begin{eqnarray}
F &=& (-1)^{n-1}\int_0^\infty\frac{d\mu_1^2}{\mu_1^2}\as(\mu_1^2)\biggl[\prod_{j\ne 1}\int_0^\infty\frac{d\mu_j^2}{\mu_j^2}\rho_s(\mu_j^2)\sum_{i_j=0}^1(-1)^{i_j}\frac{\mu_1^2}{i_j\mu_j^2+\mu_1^2}\biggr]\hat{\cl{G}}(\mu_1^2/Q^2) \;+\; I_\infty[\as]\nonumber\\
&=& (-1)^{n-1}\int_0^\infty\frac{d\mu_1^2}{\mu_1^2}\as(\mu_1^2)\biggl[\prod_{j\ne 1}\int_0^\infty\frac{d\mu_j^2}{\mu_j^2+\mu_1^2}\rho_s(\mu_j^2)\biggr]\hat{\cl{G}}(\mu_1^2/Q^2) \;+\; I_\infty[\as]\;,
\end{eqnarray}
where
\beq
\hat{\cl{G}}(\eps) = -\frac{1}{2\pi i}\mbox{Disc}\hat{\cl{F}}(-\eps)\;,
\eeq
and $I_\infty$ is some integral along a contour at infinity.

Using \rf{disprel} $n-1$ times one finds
\beq
F = \int_0^\infty\frac{d\mu^2}{\mu^2}[\as(\mu^2)]^n\hat{\cl{G}}(\mu^2/Q^2) \;+\; I_\infty[\as]\;.
\eeq

As before, we expect to recover non-perturbative contributions to $F$ from the modification in the strong coupling, $\delta\as$. The behaviour of any power correction is given by the non-analytic parts of $\hat{\cl{F}}(\eps)$, in precisely the same way as that given by $\cl{F}(\eps)$ in the single-gluon case, but the coefficient multiplying the correction is different. In the case $n = 2$, we obtain the following non-perturbative contribution to the observable $F$:
\beq
\delta F = \int_0^\infty\frac{d\mu^2}{\mu^2}\left(2\as(\mu^2)\delta\as(\mu^2)-[\delta\as(\mu^2)]^2\right)\hat\cl{G}(\mu^2/Q^2)
\eeq

We shall see that the relevant non-analytic contributions are
\beq
\cl{F}\sim a_1 \eps\log\eps
\qquad\Longrightarrow\qquad
\delta F = a_1 \frac{D_1}{Q^2}\;,
\eeq
and
\beq
\cl{F}\sim \half a_2 \eps\log^2\eps
\qquad\Longrightarrow\qquad
\delta F = a_2\frac{D_1}{Q^2}\log\frac{D_2}{Q^2}\;,
\eeq
where $D_1$ and $D_2$ are defined by:
\begin{eqnarray}
D_1 &\equiv& \int_0^\infty \frac{d\mu^2}{\mu^2}\,\mu^2\,\left(2\as(\mu^2)\delta\as(\mu^2)-[\delta\as(\mu^2)]^2\right)\;,\\
\log D_2 &\equiv& \frac{1}{D_1}\int_0^\infty \frac{d\mu^2}{\mu^2}\,\mu^2\log\mu^2\,\left(2\as(\mu^2)\delta\as(\mu^2)-[\delta\as(\mu^2)]^2\right)\;.
\end{eqnarray}

While we expect the form of $\as(\mu^2)$, and hence $D_1$ and $D_2$, to be universal, we have as yet no numerical values for them, (unlike for $\cl{A}_2$). It will be necessary therefore to extract values for $D_1$ and $D_2$, either from experimental results or from some model of the form of $\alpha_s(\mu^2)$ (of which various models have been proposed \cite{asmodel}).

\sect{DIS Structure Functions}
\label{strucfns}
We consider the deep inelastic scattering of a lepton of momentum $l$ from a nucleon of momentum $P$, with momentum transfer $q$. The main kinematic variables are $Q^2 = -q^2$, the Bjorken variable $x=Q^2/2P\cdot q$ and $y=P\cdot q/P\cdot l\simeq Q^2/x s$, $s$ being the total c.m. energy squared.

The differential cross section is 
\beq
\label{diffcs}
\frac{d^2\sigma}{dxdQ^2} = \frac{2\pi\alpha^2}{Q^4}\left\{\left[1+(1-y)^2\right] F_T(x)+2(1-y) F_L(x)\right\}
\eeq
where $F_T(x) = 2F_1(x)$ and $F_L(x) = F_2(x)/x - 2F_1(x)$ are the transverse and longitudinal structure functions, which also have a weak $Q^2$ dependence which we do not show explicitly. (For simplicity we are neglecting any contribution from weak interactions, i.e. $\mbox{Z}^0$ or $\mbox{W}^\pm$ exchange.)

In the parton model, to order $\as^0$, we have
\begin{eqnarray}
F_T(x) &=& \sum_q e_q^2 [q(x)+\bar q(x)]\;,\\
F_L(x) &=& 0
\end{eqnarray}
where $q(x)$ and $\bar q(x)$ are the quark and antiquark distributions
in the target nucleon.

\begin{figure}[ht]
\begin{center}
\epsfig{file=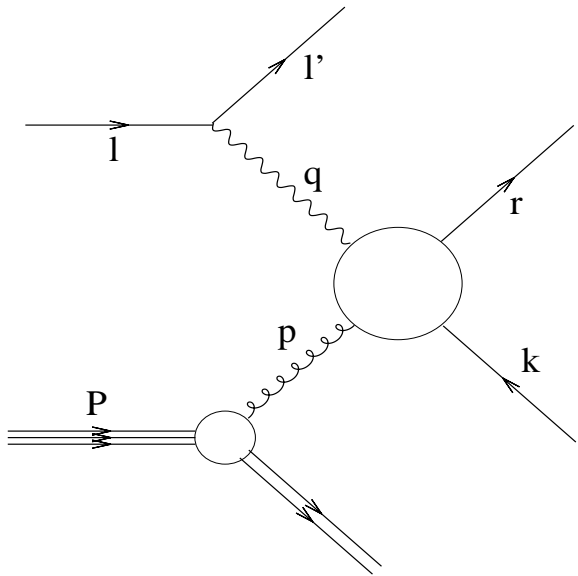, height=3.0in, width=3.0in}
\caption{\label{singfig}Flavour Singlet Contribution to Deep Inelastic Scattering}
\end{center}
\end{figure}

The $\cl{O}(\as)$ contributions, such as those shown in figure \ref{singfig}, are given by
\beq
\label{ordas}
F_i(x) = \frac{\as}{2\pi}\sum_q e_q^2 \int_x^1 \frac{d\xi}{\xi}\{C_F C_{i,q}(\xi) [q(x/\xi)+\bar q(x/\xi)] + T_R C_{i,g}(\xi)g(x/\xi)\}\;,
\eeq
where $g(x)$ is the gluon distribution, $C_F=4/3$, $T_R=1/2$. The coefficient functions $C_{i,j}(\xi)$ are the integrals in the appropriate factorisation scheme, over the final-state variable $\eta = P\cdot r/P\cdot q$ ($0\le\eta\le 1$), of \cite{PeRu}:
\begin{eqnarray}
C_{T,q}(\xi,\eta) &=& \frac{\xi^2+\eta^2}{(1-\xi)(1-\eta)}+2\xi \eta+2 \\
C_{L,q}(\xi,\eta) &=& 4\xi \eta\\
C_{T,g}(\xi,\eta) &=& \left[\xi^2+(1-\xi)^2\right]
\frac{\eta^2+(1-\eta)^2}{\eta(1-\eta)}\\
C_{L,g}(\xi,\eta) &=& 8\xi(1-\xi)\;.
\end{eqnarray}

Here we concentrate on the singlet contribution, i.e. the corrections to $C_{i,g}(\xi)$, as shown in figure \ref{singfig}.

\sect{Power Corrections in Flavour Singlet DIS}
\label{calc}
In the normal perturbative treatment of DIS, the asymptotic freedom of QCD enables us to treat the initial state partons as free particles confined within the nucleon; and so in a singlet calculation we would start from a free gluon and convolute the perturbative result with the gluon distribution function $g(x)$. We do not know how to do this in a calculation of power corrections, since the models we use consider modifications to the gluon propagator (loop insertions in the renormalon model, or, equivalently, a `mass' in the dispersive approach).
Let us therefore perform the calculation by considering our initial state gluon to be radiated from a fermionic parton. We may then try to recover the singlet contribution to the power corrections by deconvoluting the result with the quark to gluon splitting function, as performed in \cite{stein}, or we may leave the result as it is and interpret it as a second order non-singlet contribution. We might hope that these two interpretations would give similar predictions for power-suppressed corrections.

\begin{figure}[ht]
\begin{center}
\epsfig{file=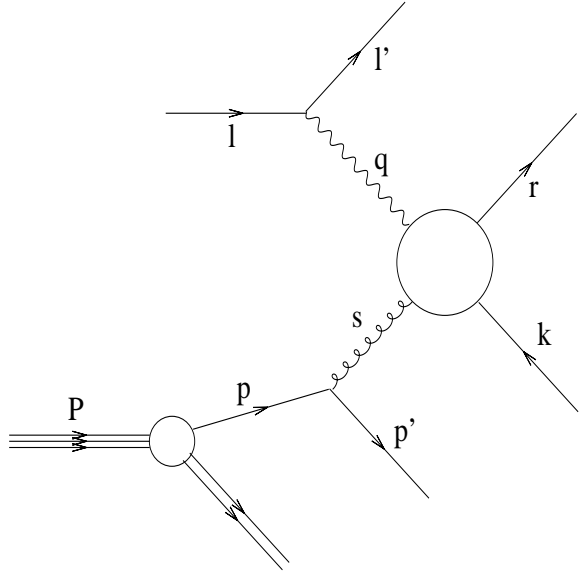, height=3.0in, width=3.0in}
\caption{\label{diags}Diagrams Generating Flavour Singlet Contribution}
\end{center}
\end{figure}

Consider the contribution to DIS from the diagrams in figure \ref{diags}. It is convenient to work in the {\it Breit frame of reference} \cite{PeRu,breit}, which is the rest-frame of $2xP+q$. In this frame the momentum transfer $q$ is purely spacelike, and we choose to align it along the $+z$ axis.

The momentum of the initial state parton is $p = xP/\xi$, $(x\le\xi\le 1)$; and let us introduce the variables $\rho = - s^2 / Q^2$, $\eta = P\cdot r/P\cdot q$, $\bar{\eta} = P\cdot k/P\cdot q$ , $\chi$ the azimuthal angle between $r$ and $s$, and $\gamma$ the azimuthal angle between $r$ and $k$. There is also an overall azimuthal angle $\phi$, the dependence on which is trivial.

In the Breit frame the kinematics are given by:
\begin{eqnarray}
P &=& \half Q (1/x,0,0,-1/x) \\
p &=& \half Q (1/\xi,0,0,-1/\xi) \\
q &=& \half Q (0,0,0,2) \\
s &=& \half Q (s_0,s_\perp\cos\chi,s_\perp\sin\chi,s_3) \\
r &=& \half Q (z_0,z_\perp,0,z_3) \\
k &=& \half Q (\bar{z}_0,\bar{z}_\perp\cos\gamma,\bar{z}_\perp\sin\gamma,\bar{z}_3)\;.
\end{eqnarray}

The definitions of $\rho$, $\eta$ and $\bar{\eta}$ along with the on-shell conditions for the outgoing particles require that
\begin{eqnarray}
s_0 = \frac{1}{\xi}-\rho\xi-\frac{s_\perp^2}{4\rho\xi}\hspace{0.3in} & &
s_3 = -\frac{1}{\xi}-\rho\xi+\frac{s_\perp^2}{4\rho\xi} \\
z_0 = \eta+\frac{z_\perp^2}{4\eta}\hspace{0.75in} & &
z_3 = \eta-\frac{z_\perp^2}{4\eta} \\
\bar{z}_0 = \bar{\eta}+\frac{\bar{z}_\perp^2}{4\bar{\eta}}\hspace{0.75in} & &
\bar{z}_3 = \bar{\eta}-\frac{\bar{z}_\perp^2}{4\bar{\eta}}\;.
\end{eqnarray}

Conservation of the 0th and 3rd components of 4-momentum give the conditions
\begin{eqnarray}
\eta + \bar{\eta} + \rho\xi &=& 1 \\
\frac{z_\perp^2}{4\eta}+\frac{\bar{z}_\perp^2}{4\bar{\eta}}+\frac{s_\perp^2}{4\rho\xi} &=& \frac{1-\xi}{\xi}\;,
\end{eqnarray}
while conservation of transverse momentum requires that $s_\perp$, $z_\perp$ and $\bar{z}_\perp$ satisfy the triangle inequalities
\begin{eqnarray}
\vert z_\perp-s_\perp\vert &\le& \bar{z}_\perp \\
\vert z_\perp-\bar{z}_\perp\vert &\le& s_\perp\;.
\end{eqnarray}

Two variables, $\alpha$ and $\beta$, are required to parametrise the permitted values of $s_\perp$, $z_\perp$ and $\bar{z}_\perp$. Let us choose to write:
\begin{eqnarray}
z_\perp^2 &=& \frac{4\alpha\eta(1-\eta)(1-\xi)}{\xi} \\
\bar{z}_\perp^2 &=& \frac{4\bar{\eta}(1-\xi)}{(1-\eta)\xi}\left[(1-\alpha)\rho\xi+\alpha\eta\bar{\eta}-2\cos\beta\sqrt{\alpha(1-\alpha)\eta\bar{\eta}\rho\xi}\right] \\
s_\perp^2 &=& \frac{4\rho(1-\xi)}{1-\eta}\left[(1-\alpha)\bar{\eta}+\alpha\eta\rho\xi+2\cos\beta\sqrt{\alpha(1-\alpha)\eta\bar{\eta}\rho\xi}\right]\;.
\end{eqnarray}

Given $s_\perp$, $z_\perp$ and $\bar{z}_\perp$, the angles $\chi$ and $\gamma$ are determined up to a sign. We may then choose $\alpha$, $\beta$, $\eta$ and $\rho$ as the independent variables, with phase space
\begin{equation}
0\le\alpha\le 1,\;0\le\beta\le\pi,\;0\le\eta\le 1,\;0\le\rho\le (1-\eta)/\xi .
\end{equation}

To estimate power corrections to perturbative calculations, we must perform the calculations as though the gluons had small non-zero masses $\mu_1^2 = \epsilon_1 Q^2$ and $\mu_2^2 = \epsilon_2 Q^2$. Using the machinery of section \ref{dispsec} we may write
\begin{equation}
{\cal F}_i(x;\epsilon_1,\epsilon_2) = \frac{\epsilon_1\hat{\cal F}_i(x;\epsilon_1)-\epsilon_2\hat{\cal F}_i(x;\epsilon_2)}{\epsilon_1-\epsilon_2}\;,
\end{equation}
where
\begin{equation}
\hat{\cal F}_i(x;\epsilon) = \frac{\alpha_s^2 T_R C_F}{(2\pi)^2} \sum_q e_q^2 \int_x^1\frac{d\xi}{\xi} C_i(\xi;\epsilon) q(x/\xi)\;.
\end{equation}

The quantities $C_i$ are the integrated perturbative matrix elements with modified gluon propagators. (Note that these are not the same quantities as those introduced in equation \rf{ordas}, since they refer to different diagrams.)

To integrate the matrix elements we apply the operator
\begin{equation}
\int\frac{d^3\b{r}}{(2\pi)^3 2r^0}\frac{d^3\b{k}}{(2\pi)^3 2k^0}\frac{d^3\b{p}^\prime}{(2\pi)^3 2p^{\prime 0}}\:(2\pi)^4\delta^4(p+q-k-r-p^\prime)\;.
\end{equation}
Integrating away $\b{p}^\prime$ and making substitutions for $\b{r}$ and $\b{k}$ gives
\begin{equation}
\frac{Q^2}{32(2\pi)^5}\int\frac{z_\perp dz_\perp d\phi d\eta}{\eta}\frac{\bar{z}_\perp d\bar{z}_\perp d\gamma d\bar{\eta}}{\bar{\eta}}\:\frac{\delta\left(A+\sqrt{B+2z_\perp\bar{z}_\perp\cos\gamma}\right)}{\sqrt{B+2z_\perp\bar{z}_\perp\cos\gamma}}\;,
\end{equation}
where $A$ and $B$ do not depend on $\gamma$.

Next we integrate over $\gamma$. There are two values satisfying the integration condition, and they differ by a sign. This gives
\begin{equation}
\frac{Q^2}{32(2\pi)^5}\int\frac{d\eta}{\eta}\frac{d\bar{\eta}}{\bar{\eta}}\frac{dz_\perp^2 d\bar{z}_\perp^2}{2z_\perp\bar{z}_\perp\vert\sin\gamma\vert}d\phi\;.
\end{equation}

Applying the parametrisation in terms of $\alpha$ and $\beta$, we find that
\begin{eqnarray}
\frac{\partial(z_\perp^2,\bar{z}_\perp^2)}{\partial(\alpha,\beta)} &=& 32(1-\xi)^2\sin\beta\sqrt{\alpha(1-\alpha)\eta^3\bar{\eta}^3\rho/\xi^3} \\
2z_\perp\bar{z}_\perp\vert\sin\gamma\vert &=& 8(1-\xi)\sin\beta\sqrt{\alpha(1-\alpha)\eta\bar{\eta}\rho/\xi}\;.
\end{eqnarray}
Therefore the integral operator is
\begin{equation}
\frac{Q^2(1-\xi)}{8(2\pi)^5}\int d\rho d\eta d\alpha d\beta d\phi\;;
\end{equation}
and hence
\begin{equation}
C_i(\xi;\epsilon) = -\frac{1}{\epsilon}\frac{Q^2(1-\xi)}{8(2\pi)^5}\int_0^{1/\xi}d\rho\int_0^{1-\rho\xi}d\eta\int_0^1d\alpha\int_0^\pi d\beta\int_0^{2\pi}d\phi\frac{f(\xi,\eta,\alpha,\beta,\phi,\rho)}{\rho+\epsilon}\;,
\end{equation}
where the function $f$ is obtained from the appropriate matrix elements, excluding the denominators of the gluon propagators.

The integration over $\phi$ is trivial.

The part of this that is nonanalytic as $\epsilon\rightarrow 0$ is the contribution to the integral near $\rho = 0$. Therefore we try to proceed by expressing $f$ as a series expansion in $\rho^{\frac{1}{2}}$ about 0, up to ${\cal O}(\rho^2)$. In order to produce valid series expansions, it transpires that we have to divide the $\eta$ integral into three regions:
\begin{enumerate}
\item $0\le\eta\le\rho/\kappa$,
\item $\rho/\kappa\le\eta\le 1-\rho\xi-\rho/\kappa$, and
\item $1-\rho\xi-\rho/\kappa\le\eta\le 1-\rho\xi$,
\end{enumerate}
where $\kappa$ is some small arbitrary quantity on which the final answer should not depend. (In practice we perform the calculation in the limit of small $\kappa$.)

For region (ii) we may directly expand in $\rho$ without any problems. The integral over $\beta$ is then straightforward, being a sum of terms of the form $\mbox{const.}\cos^n\beta$, and upon performing this integral the terms in non-integer powers of $\rho$ vanish. The integrals over $\alpha$, then $\eta$, then $\rho$ are not difficult, provided we note that the non-analytic parts of the $\rho$ integrals are given by:
\begin{eqnarray}
\int_0^c\frac{\rho^n}{\rho+\epsilon}d\rho &\to& (-1)^{n-1}\epsilon^n\log\epsilon\;, \\
\int_0^c\frac{\rho^n\log\rho}{\rho+\epsilon}d\rho &\to& {\textstyle\frac{1}{2}}(-1)^{n-1}\epsilon^n\log^2\epsilon\;.
\end{eqnarray}
Note that because of the limits we have chosen for the $\eta$ integral, contributions up to ${\cal O}(\rho^2)$ will also arise from all higher terms in the series expansion of $f$. These contributions vanish as $\kappa\rightarrow 0$, so the calculation performed in this limit is valid.

For region (i) such a naive power series becomes invalid, but we may write $\eta = \lambda\rho$ and then expand as before. In contrast to the situation in region (ii) there are now no terms in non-integer powers of $\rho$. The integration over $\beta$ may be performed by writing $t = \tan(\beta/2)$, and that over $\alpha$ by making a series of substitutions. The integration over $\lambda$ (i.e. $\eta$) can then be performed, and, since we are only concerned about the limit $\kappa\rightarrow 0$, we neglect all terms that vanish in this limit. The $\rho$ integration then proceeds as above.

For region (iii), we use the symmetry between the outgoing quark and antiquark: if we replace $\eta$ by $1-\eta-\rho\xi$ we see that the integrals over regions (i) and (iii) are equal.

Putting all this together we have:
\begin{eqnarray}
\xi C_T &=& -{\textstyle\frac{2}{9}}(2-63\xi+63\xi^2-2\xi^3+12\log\xi-27\xi\log\xi-27\xi^2\log\xi+12\xi^3\log\xi)\log\eps\nonumber\\ & & +{\textstyle\frac{2}{3}}(4+3\xi-3\xi^2-4\xi^3+6\xi\log\xi+6\xi^2\log\xi)(\log\xi-1+\half\log\eps)\log\eps\nonumber\\ & & +{\textstyle\frac{2}{5}}(2+25\xi^2-25\xi^3-2\xi^5+15\xi^2\log\xi+15\xi^3\log\xi)\eps\log\eps\nonumber\\ & & -2(5\xi^2-5\xi^3+2\xi^2\log\xi+2\xi^3\log\xi)(\log\xi-1+\half\log\eps)\eps\log\eps\;,\\
\xi C_L &=& -{\textstyle\frac{8}{3}}(1-3\xi+2\xi^3-3\xi^2\log\xi)\log\eps\nonumber\\ & & -{\textstyle\frac{8}{225}}(17+75\xi^2-125\xi^3+33\xi^5+30\log\xi+75\xi^3\log\xi-45\xi^5\log\xi)\eps\log\eps\nonumber\\ & & +{\textstyle\frac{8}{15}}(2-15\xi^2+10\xi^3+3\xi^5-15\xi^3\log\xi)(\log\xi-1+\half\log\eps)\eps\log\eps\;.
\end{eqnarray}

The terms that diverge as $\eps\to 0$ are responsible for the logarithmic scaling violations to the structure function (c.f. \cite{dmw}). To see this, note that the leading divergent pieces (i.e.~those proportional to $\half\log^2\eps$ for $C_T$ and to $\log\eps$ for $C_L$) can be expressed as a convolution of the quark to gluon splitting function with the first order gluon coefficient functions:
\begin{eqnarray}
\frac{2(4+3\xi-3\xi^2-4\xi^3+6\xi\log\xi+6\xi^2\log\xi)}{3\xi} &=& 2[\xi^2+(1-\xi)^2]\otimes\frac{1+(1-\xi)^2}{\xi}\qquad\\
\frac{8(1-3\xi+2\xi^3-3\xi^2\log\xi)}{3\xi} &=& 8\xi(1-\xi)\otimes\frac{1+(1-\xi)^2}{\xi}\;.
\end{eqnarray}
The subleading divergence in $C_T$, i.e.~that proportional to $\log\eps$, is factorisation-scheme-dependent.

The terms proportional to $\eps\log\eps$ and $\eps\log^2\eps$ give the $1/Q^2$ power corrections. These are consistent with the results in \cite{stein}. Comparison of the results presented here with those in \cite{stein} indicate that the terms containing the factor $(\log\xi-1+\half\log\eps)$ correspond to the contributions from the double renormalon pole, while those terms without this factor correspond to the contributions from the single renormalon pole. These results also suggest signs for the different contributions: the contributions in \cite{stein} from the single poles appear here with a positive sign whereas those from the double poles appear with a negative sign.

This leads to power corrections to the structure functions given by:
\beq
\delta F_i(x) = \frac{T_R C_F}{(2\pi)^2} \sum_q e_q^2 \int_x^1\frac{d\xi}{\xi} \delta C_i(\xi) q(x/\xi)\;,
\eeq
where
\begin{eqnarray}
\label{res1}
\delta C_T(\xi) &=& \frac{D_1}{Q^2}\biggl[\frac{2(2+25\xi^2-25\xi^3-2\xi^5+15\xi^2\log\xi+15\xi^3\log\xi)}{5\xi}\nonumber\\
& & \qquad-2(5\xi-5\xi^2+2\xi\log\xi+2\xi^2\log\xi)\log\frac{D_2\xi}{eQ^2}\biggr]\\
\label{res2}
\delta C_L(\xi) &=& \frac{D_1}{Q^2}\biggl[-\frac{8(17+75\xi^2-125\xi^3+33\xi^5+30\log\xi+75\xi^3\log\xi-45\xi^5\log\xi)}{225\xi}\nonumber\\
& & \qquad +\frac{8(2-15\xi^2+10\xi^3+3\xi^5-15\xi^3\log\xi)}{15\xi}\log\frac{D_2\xi}{eQ^2}\biggr]\;.
\end{eqnarray}

\begin{figure}[ht]
\begin{center}
\epsfig{file=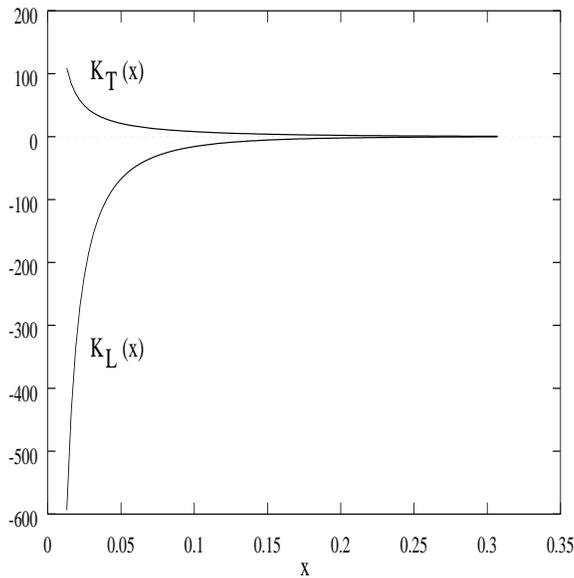, height=3.0in, width=3.0in}
\caption{\label{graph}Graph showing $K_T(x)$ and $K_L(x)$.}
\end{center}
\end{figure}

\sect{Results and Conclusions}
\label{concl}
The $1/Q^2$ power corrections arising from the diagrams shown in figure \ref{diags} are given above in equations \rf{res1} and \rf{res2}. While we do not know the values of $D_1$ and $D_2$, we can still make some qualitative predictions. Figure \ref{graph} shows plots of $K_T(x)$ and $K_L(x)$, defined by
\beq
\delta F_i(x) = \frac{D_1}{Q^2}\frac{T_R C_F}{(2\pi)^2}K_i(x)\;,
\eeq
i.e.~$K_i(x)$ are the coefficients of the $1/Q^2$ power correction, excluding the unknown factor $D_1$. These were calculated at $Q^2 = 500 \mbox{GeV}^2$, using the corresponding MRST (central gluon) parton distributions \cite{mrst}. The value of $D_2/e$ was set to be $0.06 \mbox{GeV}^2$, i.e.~approximately $\Lambda^2$, following the result of \cite{stein}. (The qualitative behaviour of the $K_i$ does not change provided we keep $D_2 \ll Q^2$.)

As can be seen from the graph, $K_T(x)$ and $K_L(x)$ both tend to zero at large $x$, and both diverge at small $x$. $K_T(x)$ is positive and $K_L(x)$ is negative, with the magnitude of $K_L$ considerably larger at small $x$ than that of $K_T$. The corresponding quantity related to $F_2/x$, which is $K_2/x=K_T+K_L$, also therefore behaves qualitatively like $K_L$.

Therefore if $D_1$ is positive, we have a negative correction to both $F_2/x$ and $F_L$. If $D_1$ is negative then the correction is positive. In either case the corrections are important at small $x$.

We do not know the sign or magnitude of $D_1$, and there is no a priori reason why it should be either positive or negative. However, since $\cl{A}_2$ is positive, we presumably have $\delta\as(\mu^2)$ being predominantly positive. We therefore might conjecture that $D_1$ would also be positive.

\section*{Acknowledgements}
The author wishes to thank B.R.~Webber and Yu.L.~Dokshitzer for advice and helpful comments.

\end{document}